\documentclass[12pt]{article}
\usepackage[dvips]{graphicx}
\usepackage{graphicx}
\usepackage{amssymb}
\usepackage{amsmath}
\usepackage{amsfonts}

\def\hybrid{\topmargin 0pt    \oddsidemargin 0pt \evensidemargin 0pt
        \headheight 0pt
        \textwidth 6.25in       
        \textheight 23.5cm           
        \marginparwidth .875in
       \parskip 5pt plus 1pt            
        \jot = 1.5ex}
\hybrid

\newcommand{\lambdabar}{{\hbox{$\lambda_e$\kern-1.9ex\raise+0.45ex\hbox{--}
\kern+0.2ex}}}

\numberwithin{equation}{section} \numberwithin{table}{section}

\newcommand{\beq}{\begin{equation}}
\newcommand{\eeq}{\end{equation}}
\newcommand{\bi}{\begin{itemize}}
\newcommand{\ei}{\end{itemize}}
\newcommand{\bea}{\begin{eqnarray}}
\newcommand{\eea}{\end{eqnarray}}
\newcommand{\ba}{\begin{array}}
\newcommand{\ea}{\end{array}}
\newcommand{\bt}{\begin{tabular}}
\newcommand{\et}{\end{tabular}}
\newcommand{\bc}{\begin{center}}
\newcommand{\ec}{\end{center}}
\newcommand{\bdisp}{\begin{displaymath}}
\newcommand{\edisp}{\end{displaymath}}
\newcommand{\ax}{\alpha}
\newcommand{\bx}{\beta}

\newcommand{\dx}{\delta}

\newcommand{\lx}{\lambda}
\newcommand{\lb}{\bar{\lambda}}
\newcommand{\sx}{\sigma}

\newcommand{\gx}{\gamma}
\newcommand{\Px}{\Phi}
\newcommand{\tx}{\theta}

\newcommand{\tb}{\bar{\theta}}

\newcommand{\Sx}{\Sigma}
\newcommand{\Lx}{\Lambda}

\newcommand{\cL}{\mathcal{L}}

\newcommand{\RE}{\textrm{Re} \,}
\newcommand{\IM}{\textrm{Im} \,}
\newcommand{\M}{\mu}

\newcommand{\nn}{\nonumber}
\newcommand{\U}{\hat{K}}

\begin{document}

\allowdisplaybreaks[1]
\begin{titlepage}
\begin{center}

\hfill {\small hep-th/0702211}\\
\hfill {\small ZMP-HH/06-20}
\vskip 0.3cm

\vskip 1.5cm

{\large \bf Couplings of $N=1$ chiral spinor multiplets}\footnote{%
Work supported by: The German Science Foundation (DFG) and
European RTN Program  MRTN-CT-2004-503369.}\\

\vskip 1.5cm

{\bf Jan Louis$^{a,b}$  and Jacek Swiebodzinski$^{c}$ }  \\
\vskip 0.8cm

{}$^{a}${\em II. Institut f{\"u}r Theoretische Physik,
Universit{\"a}t Hamburg\\
Luruper Chaussee 149,
 D-22761 Hamburg, Germany}\\
{\tt jan.louis@desy.de}  \\

\vskip 0.4cm

{}$^{b}${\em Zentrum f\"ur Mathematische Physik,
Universit\"at Hamburg\\
Bundesstrasse 55, D-20146 Hamburg, Germany}
\vskip 0.4cm

{}$^{c}${\em I. Institut f\"ur Theoretische Physik,
Universit{\"a}t Hamburg\\
Jungiusstrasse 9, D-20355 Hamburg, Germany}\\
{\tt jswiebod@physnet.uni-hamburg.de}
\end{center}

\vskip 2cm

\begin{center} {\bf ABSTRACT } \end{center}

We derive the action for $n_L \geq 1$ chiral spinor
multiplets coupled to vector and scalar multiplets. We give the
component form of the action, which contains gauge invariant mass
terms for the antisymmetric tensors in the spinor superfield and
additional Green-Schwarz
couplings to vector fields. We observe that supersymmetry provides
mass terms for the scalars in the spinor multiplet which do not arise from
eliminating an auxiliary field. We construct the dual action by
explicitly performing the duality transformations in superspace and
give its component form.

\vfill

\noindent February 2007

\end{titlepage}

\section{Introduction}
\label{S1}

Antisymmetric tensor fields $B_{mn}$ naturally appear in the light sector of
all string theories. In four space-time  dimensions ($D=4$) massless
antisymmetric tensors are dual to
scalar fields while massive tensors are dual to massive vectors.
Therefore in the low energy effective action one has the choice to
represent these degrees of freedom in either of two dual
representations. Depending on the context one formulation
might be more convenient than the other and for this reason
both formulations have generically been developed.

Recently compactification with background fluxes and/or
compactifications on generalized geometries have been studied in
detail \cite{grana}. One novelty in these compactifications
is the appearance of massive antisymmetric tensors
\cite{Louis-Micu}. As a consequence their description in terms of
appropriate supergravities has been worked out
\cite{DSV}--\cite{SW}. In particular in
$N=1$ compactifications of type IIB on
Calabi-Yau orientifolds with $O5$- or $O9$- planes a
massive antisymmetric tensor appears when both electric and magnetic three-form
fluxes are turned on \cite{Grimm-Louis-IIB}.
The corresponding  $N=1$ superspace action was constructed in
ref.~\cite{Louis-W}.
Orientifolds of generalized geometries as discussed, for example,  in
refs.~\cite{GLW,iman} can feature more than one antisymmetric tensor.
Therefore it is of interest to generalize the analysis of
\cite{Louis-W} and discuss the couplings of a set of $n_L$ massive
antisymmetric tensors to vector and chiral multiplets. This is the
purpose of the present paper.

In $N=1$ supersymmetry the three form field strength of the
antisymmetric tensor is part of a linear multiplet $L$ \cite{FWZ}--\cite{derendinger-1}. The
antisymmetric tensor itself resides in the chiral spinor multiplet
$\Phi_\ax$.
Whenever the antisymmetric tensor is massless the
supersymmetric action  is described in
terms of $L$ only. Any mass term for $B_{mn}$ destroys the two-form
gauge invariance. However, with the help of appropriate couplings to
vector fields gauge invariance can be restored.
The resulting Lagrangian is of the St\"uckelberg type \cite{Stueck}
where the vector fields provide the `longitudinal' degrees of freedom
to render $B_{mn}$ massive. Put differently, in a unitary gauge the
antisymmetric tensor `eats' a vector field and becomes massive.
A similar mechanism can be employed for $n_L$ antisymmetric tensors
as long as enough ($n_V\geq n_L$) vector fields are coupled.
Therefore the first goal of this paper is the derivation of
a $N=1$ superspace action for $n_L$
chiral spinor multiplets $\Phi_\ax^I, I= 1,\ldots n_L$  coupled to
$n_V$ vector multiplets $V^A, A=1,\ldots n_V$. Furthermore
the gauge couplings of the vector multiplets are allowed
to depend on $n_C$  chiral multiplets $N^i$, $i=1,\dots, n_C$.

As we already stated a massless antisymmetric tensor is dual to a
scalar while a massive one is dual to a massive vector. This
duality is also manifest at the level of superfields where a
linear multiplet is dual to a chiral multiplet while a massive
spinor multiplet is dual to a massive vector multiplet. Thus  our
second aim is to construct the dual theory in superspace.

This paper is organized as follows. In section \ref{S2} we
introduce the notions of the linear and the chiral
spinor multiplet. By means of the St\"uckelberg mechanism we
construct the most general gauge invariant action for $n_L$
massive spinor multiplets and give its
corresponding component form. We discuss the resulting scalar
potential, which has not the standard $N=1$ form due to a
contribution from the chiral spinor multiplet.
In section \ref{S3} we perform the duality
transformations and rewrite the action in terms of $n_V-n_L$
massless and $n_L$ massive vector multiplets. Finally in the appendix
we present the supersymmetry transformations of the
chiral spinor multiplet and give a modification of these
transformations which preserves the WZ-gauge. This allows us to
discuss the order parameters for supersymmetry breaking.

\section{Spinor superfields coupled to vector and chiral multiplets}
\label{S2}

In $N=1$ supersymmetry an antisymmetric tensor $B_{mn}$ is part of
a chiral spinor superfield $\Phi_\ax$ while its three-form field
strength $H_{mnp}$ resides in a linear multiplet $L$. In this
section we consider a set of $n_L$ linear multiplets $L^I$
and the corresponding $n_L$ chiral spinor multiplets
$\Phi_\ax^I, \, I=1, \dots, n_L$. We review some of their
properties and construct a gauge invariant action.

The linear multiplet is a real superfield, defined by the
constraint \cite{siegel}
\beq\label{L}
 D^2 L^I = \bar  D^2 L^I =0 \ ,
\eeq
where $D_\ax$ is the superspace covariant
derivative.\footnote{Throughout the paper we are using the conventions of
  ref.~\cite{WB}.} The
$\theta$-expansion of $L^I$ reads \beq \label{nL-linearsuperfield}
L^{I}=C^{I}+\theta\eta^{I}+\bar{\theta}\bar{\eta}^{I}+ \tfrac12
\theta\sigma^{m}\bar{\theta}\epsilon_{mnpq}H^{npq \, I} -
\tfrac{i}{2}(\theta\theta)\bar{\theta}\bar{\sigma}^{m}\partial_{m}\eta^{I}-
\tfrac{i}{2}(\bar{\theta}\bar{\theta})\theta\sigma^{m}\partial_{m}\bar{\eta}^{I}-
\tfrac{1}{4}\theta\theta\bar{\theta}\bar{\theta}\square C^{I} \ .
\eeq
Here $C^I$ are real scalars, $\eta^I$ are Weyl fermions and
$H^I_{mnp}=\partial_{[m}B^I_{np]}$ are the field strengths of the
antisymmetric tensors $B^I_{np}$.

Each antisymmetric tensor $B^I_{mn}$ is contained in a chiral
 spinor superfield $\Phi^I_\ax$ defined by \cite{siegel}
\beq\label{LdefPhi} L^I= \tfrac12 ( {{D}} ^ \alpha \Phi^I _\alpha
+\bar {{D}} _ {\dot \alpha} \bar \Phi ^{\dot \alpha\, I}),\qquad
\bar {{D}} _{\dot \beta} \Phi^I _ \alpha = 0 \ .
 \eeq
The
$\Phi^I_\ax$ enjoy the $\theta$-expansion
\begin{equation}\label{entphi}
\begin{split}
\Phi^I_{\alpha}=&\chi^I_{\alpha} -
\theta_{\gamma}\Big(\tfrac12 {\delta_{\alpha}}^{\gamma}(C^I+iE^I)
+
\tfrac{1}{4}{(\sigma^m\bar{\sigma}^n)_{\alpha}}^{\gamma}B^I_{mn}\Big)
+ \theta\theta\left(\eta^I_{\alpha} +
i{\sigma_{\alpha\dot{\alpha}}}^m\partial_m\bar{\chi}^{\dot{\alpha}I}\right) \ ,
\end{split}
\end{equation}
where $\chi^I_{\alpha}$ are additional Weyl fermions and $E^I$
additional scalars.
Due to its definition \eqref{LdefPhi} the $L^I$ are invariant under the
gauge transformations
 \beq
\label{nL-eichtrPhi} \Phi^I_{\ax} \longrightarrow \Phi^I_{\ax} +
\tfrac{i}{8} \bar{D}^2 D_{\ax}\Lambda^I \ , \eeq
where the $\Lx^I$ are
real superfields. The expressions $\bar {{D}}^2 {{D}}_\alpha
 \Lambda^I$ are chiral and we therefore can write\footnote{The expansion
 has the same structure as the field strength of the vector multiplet
which we introduce in (\ref{Wexpansion}).
   To avoid confusions with (\ref{Wexpansion}) we have
 hatted the corresponding component fields of (\ref{eichtermPhi}).}
 \beq \label{eichtermPhi}
\tfrac{i}{8} \bar {D}^2  {D}_ \alpha \Lambda^I = -\tfrac12
\hat\lambda^{I}_\alpha -\left({\delta_\alpha}^\gamma
\tfrac{i}{2}\, \hat D ^{I} +\tfrac14 {(\sigma^m
\sigma^n)_\alpha}^\gamma
\left(\partial_m\Lambda_n^{I}-\partial_n\Lambda_n^{I}\right)\right)\theta_\gamma
-\tfrac{i}{2}\theta\theta
n\sigma_{\alpha\dot{\alpha}}^m\partial_m\bar{ \hat \lambda} ^{ \,
\dot {\alpha} I} \ . \eeq We immediately see that the fields
$\chi^I_\ax$ and $E^I$ defined in the $\theta$-expansion of the
superfield $\Phi^I_\ax$ in \eqref{entphi} can be gauged away by
$\hat\lambda^{I}_\alpha$ and $\hat D ^{I}$ using (\ref{nL-eichtrPhi}). This
leaves only the physical degrees of freedom $C^I$, $B^I_{mn}$ and
$\eta^I$ in the component expansion of $\Phi^I_\ax$. Thus in this
WZ-gauge we have
\begin{equation}\label{Phiexpansion1}
\begin{split}
\Phi^I_{\alpha}=&-\theta_{\gamma}
\left(\tfrac12 {\delta_{\alpha}}^{\gamma}C^I
+\tfrac{1}{4}{(\sigma^m\bar{\sigma}^n)_{\alpha}}^{\gamma}B^I_{mn}\right)
+\theta\theta\eta^I_{\alpha}
\ ,
\end{split}
\end{equation}
and the left-over gauge invariance is the standard two-form gauge
invariance \bea \label{transkompvonPhi2} B^I_{mn} &\to&
B^I_{mn}+\partial_m\Lambda^I_n-\partial_n\Lambda^I_m \ ,\qquad C^I
\to C^I \ ,\qquad\ \eta^I_{\alpha} \to \eta^I_{\alpha}\ .\eea

The superfields $\Phi^I_\ax$ and $L^I$ can be used to construct a
gauge invariant action. The
kinetic term is given by
\begin{equation} \label{nL-Lkin} \mathcal{L}_{kin}= -
\int d^2\theta d^2\bar{\theta}K( L^I ) \ ,
\end{equation}
where $K( L^I )$ is an arbitrary real function of the $L^I$. In
components (\ref{nL-Lkin}) reads
\beq
\begin{split}
\label{nL-Lkin-comp} \mathcal{L}_{kin} =  -
&\tfrac{1}{4} K_{IJ} \Big((\partial_mC^J)(\partial^mC^I) +
i(\eta^{I}\sigma^{m}\partial_{m}\bar{\eta}^{J}+
\bar{\eta}^{I}\bar{\sigma}^{m}\partial_{m}\eta^{J})+ \tfrac32
H^{mnp \, I}H_{mnp}^{I}\Big) \\  - &\tfrac{1}{8} K_{IJK}
\left(\eta^{K}\sigma^{m}\bar{\eta}^{I}\epsilon_{mnpq}H^{npq \,
J}\right) - \tfrac{1}{4!} K_{IJKL}
\left(\tfrac{3}{2}\eta^{I}\eta^{J}\bar{\eta}^{K}\bar{\eta}^{L}\right)
\ ,
\end{split} \eeq where we abbreviated
\beq
K_{IJ \ldots K} \equiv \frac{\partial^{n} K(C)}{\partial
C^{I}\partial C^{J}\ldots \partial C^{K}} 
\ .
\eeq


In addition to the kinetic term we can add a mass term for the
$B^I_{mn}$ if we introduce a set of Abelian vector  multiplets
$V^A, A=1, \dots, n_V$. As we will see they can be used to ensure
the gauge invariance \eqref{transkompvonPhi2} and they also
provide the necessary degrees of freedom in order to render the
$B^I_{mn}$ massive. Let us denote the field strengths of the
vector multiplets by $W^A_\ax = - \tfrac14 \bar D^2 D_\ax V^A$
with the component expansion \beq\label{Wexpansion}
W_\alpha^A=-i\lambda^A_\alpha+\Big({\delta_\alpha}^\beta D^A -
\tfrac{i}{2} {(\sigma^m\bar\sigma^n)_\alpha}^\beta F^A_{mn}\Big)
\theta_\beta +
\theta\theta\sigma_{\alpha\dot\alpha}^m\partial_m\bar\lambda^{\dot{\alpha}A}\
. \eeq Here $F^A_{mn}=\partial_m v_n^A -\partial_n v_m^A$ are the
field strengths of $n_V$ $U(1)$ gauge bosons
$v_n^A$.\footnote{$W_\alpha^A$ is invariant under the standard
$U(1)$ gauge
  invariance $V^A \to V^A + \Sx^A + \bar{\Sx}^A$ where $\Sx^A$ are
  chiral superfields.}
The linear
combination \beq \label{invkomb-VL}
2im_{\phantom{A}J}^{A}\Phi^J_{\bx} - W^A_{\bx} \  \eeq is gauge
invariant under \eqref{nL-eichtrPhi}
provided we assign the following transformation laws to
the $V^A$
\beq \label{trsuplang}
\begin{split}
 V^A \to V^A + m_{\phantom{A}J}^{A} \Lambda^J \ , \qquad
W^A_{\bx} \to & W^A_{\bx} - \tfrac{1}{4} m_{\phantom{A}J}^{A}\bar{D}^2
 D_{\bx}\Lambda^J \ .
\end{split}
\eeq In (\ref{invkomb-VL}) and (\ref{trsuplang}) we have
introduced the constant coupling matrix $m_{\phantom{A}J}^{A}$
which we demand to be real. The linear combination
(\ref{invkomb-VL}) can be used to build (Lorentz and gauge
invariant) mass-terms for $\Phi^J_{\bx}$.
 However this is
not the only possible gauge invariant term. Permitting the
Lagrangian to be invariant only up to a total derivative we can
also add the term $2\int d^2 \tx
e_{AI}\Px^I\left(W^A-im^A_{\phantom{A}J}\Px^J\right) +\rm{ h.c.}$
where $e_{AI}$ is a constant real matrix. Note that in this expression
only the symmetric part of the product $e_{AI}m^A_{\phantom{A}J}$ appears.
The gauge invariance of this additional term can be most easily seen
by first rewriting the term as
\beq \label{ginvrew}
-2\int d^4\tx e_{AI}L^IV^A - \Big( i \int d^2 \tx
e_{AI}m^A_{\phantom{A}J} \Px^I \Px^J + \textrm{ h.c.} \Big) \ ,
\eeq where we used $d^2\tx = - \tfrac14 D^2$, (\ref{LdefPhi}) and the
definition of $W_\alpha$. Using
 (\ref{nL-eichtrPhi}) and (\ref{trsuplang}) we perform the gauge
 transformations on (\ref{ginvrew}). For the first term we obtain
 \beq \label{dxa} \dx \int d^4 \tx (-2 e_{AI} L^I V^A) = -2 \int d^4 \tx
 e_{AI}m^A_{\phantom{A}J} L^I
 \Lx^J\eeq
as $\dx L^I = 0$. Transformation of the second term reads
\beq \label{dxb}
\begin{aligned}
\dx \int d^2 \tx \left(- i
e_{AI}m^A_{\phantom{A}J} \Px^I \Px^J \right) + \textrm{ h.c.} &=
- i\int d^2 \tx e_{AI}m^A_{\phantom{A}J} \left( \tfrac{i}{4}
\Phi^I \bar D^2 D
\Lx^J\right) + \textrm{ h.c.}  \\
&= - \int d^4 \tx e_{AI}m^A_{\phantom{A}J} \left(  \Phi^I (D
\Lx^J) + \bar \Phi^I (\bar D \Lx^J) \right) \\ 
&= 2 \int d^4 \tx e_{AI}m^A_{\phantom{A}J} \Lx^J L^I +
\textrm{total derivative} \ , \phantom{-} 
\end{aligned}
\eeq 
where (\ref{LdefPhi}) and the
chirality of $\Phi$ was used. In (\ref{dxb}) again
only the symmetric part of $e_{AI}m^A_{\phantom{A}J}$ enters while the
variation in  (\ref{dxa}) contains also the antisymmetric part. Therefore
gauge invariance of (\ref{ginvrew}) requires to impose the condition
$e_{AI}m^A_{\phantom{A}J}=e_{AJ}m^A_{\phantom{A}I}$.\footnote{We thank
U.~Theis for discussions on this point.}
Thus the most general gauge invariant
action of $n_L$ massive spinor multiplets coupled to $n_V$ vector
multiplets is given by \beq \label{Lm-VL} \cL_m =
 \tfrac{1}{4}\int d^2\tx \Big( f_{AB}
 (2im^A_{\phantom{A}I}\Px^I-W^A)(2im^B_{\phantom{B}J}\Px^J-W^B)
 + 2e_{AI}\Px^I\left(W^A-im^A_{\phantom{A}J}\Px^J\right) \Big) +
 \rm{ h.c.} \ .
\eeq
The matrix $f_{AB}$ is the gauge coupling function of the vector
multiplets which can  depend
holomorphically on additional chiral multiplets which we denote by
$N^i, i=1,\ldots, n_C$.\footnote{Of course we also need to add kinetic
  terms for the $N^i$ but since they play no role here we omit them in
  the following.}
The Lagrangian \eqref{Lm-VL} is our first result which coincides
with the Lagrangian of ref.~\cite{Louis-W} in the limit of
one linear multiplet and was also given previously in ref. \cite{theis}.

In components the  Lagrangian (\ref{Lm-VL}) reads
\bea\label{lmkomp}
\mathcal{L}_{\rm m}
 &= &-\tfrac{1}{4}\mathrm{Re}f_{AB}\check{F}^A_{mn}\check{F}^{B\,mn}
+\tfrac{1}{8}\mathrm{Im}f_{AB}\varepsilon^{klmn}\check{F}^A_{kl}\check{F}^B_{mn}
-\tfrac{1}{16}\epsilon^{klmn}e_{AI}B^I_{kl}(\check{F}^A_{mn}+F^A_{mn})
\nn \\
&&+\tfrac12  \mathrm{Re}f_{AB} D^A D^B -\tfrac12 (e_{AI}
+2\mathrm{Im}f_{AB}m^B_{\phantom{B}I})C^ID^A - \tfrac12
\mathrm{Re}f_{AB} m^A_{\phantom{A}I}m^B_{\phantom{B}J} C^IC^J\nn\\
&&-\tfrac{i}{2}\,f_{AB}\lambda^A\sigma^m\partial_m\bar{\lambda}^B
-\tfrac{i}{2}\,\bar f_{AB}\bar{\lambda}^A\bar{\sigma}^m\partial_m{\lambda}^B\nn\\
&&-\tfrac12 (ie_{AI} +2 f_{AB}m^B_{\phantom{B}I})
\eta^I\lambda^A-\tfrac12(-ie_{AI} +2
\bar f_{AB}m^B_{\phantom{B}I})\bar{\eta}^I\bar{\lambda}^A\\
&&-\tfrac{1}{2\sqrt{2}}\,\partial_i
f_{AB}(m^A_{\phantom{A}I}C^I-iD^A)\chi^i\lambda^B
-\tfrac{1}{2\sqrt{2}}\,\partial_{\bar i} \bar f_{AB}(m^A_{\phantom{A}I}C^I+iD^A)
\bar{\chi}^{\bar i}\bar{\lambda}^B\nn\\
  &&-\tfrac{1}{2\sqrt{2}}\,\partial_i
 f_{AB}\check{F}^A_{mn}\chi^i\sigma^{mn}\lambda^B
 -\tfrac{1}{2\sqrt{2}}\,
 \partial_{\bar i}\bar f_{AB}\check{F}^A_{mn}\bar{\chi}^{\bar i}\bar{\sigma}^{mn}\bar{\lambda}^B\nn\\
  &&-\tfrac{1}{4}F^i\partial_i f_{AB}\lambda^A\lambda^B
-\tfrac{1}{4}\bar F^{\bar i}\partial_{\bar i}\bar
f_{AB}\bar{\lambda}^A\bar{\lambda}^B\nn
+\tfrac{1}{8}\chi^i\chi^l\partial_i\partial_lf_{AB}\lambda^A\lambda^B
+\tfrac{1}{8}\bar{\chi}^{\bar i}\bar{\chi}^{\bar l}\partial_{\bar i}\partial_{\bar
 l}\bar f_{AB}
\bar{\lambda}^A\bar{\lambda}^B\ ,
\eea
where we defined \beq \label{fcheck} \check{F}_{mn}^A \equiv
F^A_{mn}-m^A_{\phantom{A}I}B^I_{mn} \ , \eeq and used as the component
expansion of $N^i$
\bea
N^i&=&A^i+\sqrt{2}\theta\chi^i+\theta\theta F^i\ ,\\
f_{AB}(N)&=&f_{AB}(A)+\sqrt{2}\theta\chi^i\partial_if_{AB}(A)
+\theta\theta\big(F^i\partial_if_{AB}(A)
-\tfrac{1}{2}\chi^i\chi^j\partial_i\partial_jf_{AB}(A)\big)\ .
\nonumber \eea
(We abbreviate $\partial_i \equiv \frac{\partial}{\partial A^i}$.)
 The auxiliary fields $D^A$ may be eliminated by their
equations of motion
 \beq \label{eom-DA} D^A =\tfrac{1}{2}(\RE f)^{-1AB}\Big( (e_{BI}
+2\mathrm{Im}f_{BC}m^C_{\phantom{C}I})C^I
-\tfrac{i}{\sqrt{2}}(\partial_i f_{BC}\chi^i\lambda^C
-\partial_{\bar i}\bar f_{BC}\bar{\chi}^{\bar i}\bar{\lambda}^C)
\Big) \ . \eeq Inserting (\ref{eom-DA}) into (\ref{lmkomp}) we
obtain
\bea \label{lmkompohneHF}
\mathcal{L}_{\rm m} &=&
-\tfrac{1}{4}\mathrm{Re}f_{AB}\check{F}^A_{mn}\check{F}^{B\,mn}
+\tfrac{1}{8}\mathrm{Im}f_{AB}\varepsilon^{klmn}\check{F}^A_{kl}\check{F}^B_{mn}
-\tfrac{1}{16}\epsilon^{klmn}e_{AI}B^I_{kl}(\check{F}^A_{mn}+F^A_{mn})
- V\nn
\\
&&-\tfrac12 (ie_{AI} +2 f_{AB}m^B_{\phantom{B}I})
\eta^I\lambda^A-\tfrac12(-ie_{AI} +2
\bar f_{AB}m^B_{\phantom{B}I})\bar{\eta}^I\bar{\lambda}^A\nn\\
&&-\tfrac{i}{2}\,f_{AB}\lambda^A\sigma^k\partial_k\bar{\lambda}^B
-\tfrac{i}{2}\, \bar f_{AB}\bar{\lambda}^A\bar{\sigma}^k\partial_k{\lambda}^B\nn\\
&&-\tfrac{1}{2\sqrt{2}}\partial_kf_{AB}m^A_{\phantom{A}I}C^I\chi^k\lx^B
-\tfrac{1}{2\sqrt{2}}\partial_{\bar k}\bar f_{AB}m^A_{\phantom{A}I}C^I
\bar{\chi}^{\bar k}\lb^B\nn\\
&&+\tfrac{i}{4\sqrt{2}}(\RE f)^{-1AB}\left(\partial_l
f_{AG}\chi^l\lx^G - \partial_{\bar l}\bar f_{AG}\bar{\chi}^{\bar
l}\lb^G \right)\left(e_{BI}
+2\mathrm{Im}f_{BC}m^C_{\phantom{C}I}\right)C^I
\\
&& + \tfrac{1}{16}(\RE f)^{-1AB}\partial_kf_{BC}\left(
\partial_lf_{AG} \chi^l \lx^G -\partial_{\bar l}\bar f_{AG}\bar{\chi}^{\bar l}\lb^G
\right)\chi^k \lx^C\nn \\ 
&&+\tfrac{1}{16}(\RE f)^{-1AB}\partial_{\bar
k}\bar f_{BC}\left(
\partial_{\bar l}\bar f_{AG} \bar{\chi}^{\bar l} \lb^G -\partial_{l}f_{AG}\chi^l\lx^G
\right)\bar{\chi}^{\bar k} \lb^C\nn \\
 &&-\tfrac{1}{2\sqrt{2}}\,\partial_k
 f_{AB}\check{F}^A_{mn}\chi^k\sigma^{mn}\lambda^B
 -\tfrac{1}{2\sqrt{2}}\,
 \partial_{\bar k}\bar f_{AB}\check{F}^A_{mn}\bar{\chi}^{\bar k}
 \bar{\sigma}^{mn}\bar{\lambda}^B\nn\\
  &&-\tfrac{1}{4}F^k\partial_k f_{AB}\lambda^A\lambda^B
-\tfrac{1}{4}\bar F^{\bar k}\partial_{\bar k}\bar f_{AB}\bar{\lambda}^A\bar{\lambda}^B\nn\\
&&  +\tfrac{1}{8}\chi^k\chi^l\partial_k\partial_lf_{AB}\lambda^A\lambda^B
+\tfrac{1}{8}\bar{\chi}^{\bar k}\bar{\chi}^{\bar l}\partial_{\bar
k}\partial_{\bar l}\bar f_{AB} \bar{\lambda}^A\bar{\lambda}^B\ ,\nn
\eea
where the scalar potential is given by \beq
\label{skalares-Potential}
\begin{split}
V
 =\tfrac{1}{8}\Big(\RE f)^{-1AB}(e_{AI}
 +2\mathrm{Im}f_{AC}m^C_{\phantom{C}I})(e_{BJ}
 +2\mathrm{Im}f_{BD}m^D_{\phantom{D}J})+ 4\RE
 f_{AB}m^A_{\phantom{A}I}m^B_{\phantom{B}J}\Big) C^IC^J \ .
\end{split}
\eeq
In order to make the contribution from the $D$-terms manifest we can
alternatively write the potential as
\beq\label{skalares-Potential-C}
V =\tfrac12 \RE f_{AB}D^AD^B + \tfrac12 \RE f_{AB}
m^A_{\phantom{A}I}m^B_{\phantom{B}J} C^IC^J
\eeq
for
$D^A=
\tfrac{1}{2}(\RE f)^{-1AB} (e_{BI}
+2\mathrm{Im}f_{BC}m^C_{\phantom{C}I})C^I$.
We see  that there is a contribution to the mass terms for the scalars
$C^I$ which does not arise from eliminating an auxiliary field.

For generic charges $e_{BJ},m^D_{\phantom{D}J}$ (i.e.\ non-zero)
the minimum of $V$ is at $C^I=0$. This follows from the fact that
$\RE f_{AB}$ is the gauge kinetic function and therefore positive definite.
As a consequence both terms in \eqref{skalares-Potential} are
manifestly positive.

To close our discussion of the Lagrangian (\ref{Lm-VL}) let us explicitly
display the mass terms for the $B^I_{mn}$. Using (\ref{fcheck})
we can write \beq \label{M2undM2T}
\begin{split}
\cL_m^{B2} =& - \tfrac{1}{4} (M^2)_{IJ} B^{I \, mn} B^J_{mn} +
\tfrac{1}{8} (M^2_T)_{IJ} \varepsilon^{klmn}B^I_{kl}B^J_{mn} \ , \\
(M^2)_{IJ}=&\phantom{-} \RE
f_{AB}m^A_{\phantom{A}I}m^B_{\phantom{B}J} \ ,
\\ (M^{2}_T)_{IJ}=& \phantom{-} \IM f_{AB}m^A_{\phantom{A}I}m^B_{\phantom{B}J}
+\tfrac{1}{2}e_{AI}m^A_{\phantom{A}J} \ .
\end{split}
\eeq
As we see the action contains an ordinary mass term $M^2$ as well as  a
topological mass term $M^{2}_T$. For
$m^A_{\phantom{A}I}=0$ both mass terms vanish and a massless
antisymmetric tensor with a Green-Schwarz coupling of the form
$e_{AI} \epsilon^{mnpq}F^A_{mn}B^I_{pq}$ is left.

\section{Dual Formulation}
\label{S3}
So far  we discussed the possible couplings of a set of spinor
superfields to Abelian vector and chiral multiplets.
In components this led to massive antisymmetric
tensors possibly with  additional Green-Schwarz couplings.
It is well known that
theories with antisymmetric tensors have an equivalent dual
formulation: a massive antisymmetric tensor is dual to a massive
vector while a massless antisymmetric tensor is dual to a scalar.
The purpose of  this section is to derive the dual of the theories
discussed in the previous section.
More specifically, we
perform a duality transformation in superfields and then expand
the dual action in components. As a warm-up we first consider the massless case
with non-trivial Green-Schwarz couplings ($m^A_{\phantom{A}I}= 0,\,
e_{AI}\neq 0$)  and then turn to the general case where also
$m^A_{\phantom{A}I} \neq  0$.

\subsection{Massless tensors with Green-Schwarz couplings}

For $m^A_{\phantom{A}I}= 0$ the action  given by
(\ref{Lm-VL}) and (\ref{nL-Lkin})
can be rewritten as
 \beq \label{actionre}
\cL = - \int d^4 \tx \left(K(L^I) + e_{AI}L^I V^A\right) +
\tfrac14 \Big( \int d^2 \tx f_{AB}W^AW^B + \textrm{ h.c.} \Big) \
, \eeq where we partially integrated using the definition of
$W_\alpha^A$, \eqref{LdefPhi} and $d^2\bar\theta = - \frac14 \bar
D^2$. We see that the entire action is expressed in terms of
linear multiplets only and no mass term for the antisymmetric
tensors is present. The Lagrangian \eqref{actionre} can be derived
from the following first-order Lagrangian \beq
\label{firstordermasselos} \cL_{first} = - \int d^4 \theta \big(
K(V^{0I}) + e_{AI}V^{0I}V^A +  V^{0I}(S_I + \bar{S}_I) \big) +
\tfrac14 \Big( \int d^2 \tx f_{AB}W^AW^B + \textrm{ h.c. } \Big) \
, \eeq where $V^{0I}$ denote $n_L$ real vector (but not linear)
superfields and $S_I$ are $n_L$ chiral superfields. Eliminating
the $S_I$ by their equations of motion we find
\bea \label{Vlinear} \bar{D}^2 V^{0I} = D^2 V^{0I} = 0 \ , \eea
where we used that a chiral $S_I$ can always be written in terms
of an unconstrained superfield $X_I$ via $S_I = \bar D^2 X_I$.
{}From \eqref{Vlinear} we learn that  $V^{0I}$ is constrained to
be a linear superfield and thus can be  identified as \beq
V^{0I}=L^I \ . \eeq Inserted back into (\ref{firstordermasselos})
using $\int d^4\tx L^I(S_I + \bar S_I) = 0$ we finally arrive at
(\ref{actionre}).

If we eliminate the $V^{0I}$ instead we obtain the dual theory in
terms of the chiral multiplets $S_I$. The equation of motion for
$V^{0I}$ reads
\beq
\label{legendrev}
\partial_{V^{0K}}K(V^{0I}) + e_{AK}V^A + S_K + \bar{S}_K
= 0 \ .
\eeq
With the help of \eqref{legendrev} we can express $V^{0K}$ as a
function of $e_{AK}V^A + S_K + \bar S_K$ and possibly of the other
$V^{0I}, \ I\neq K$.  Let us denote this function by $h^K$,
i.e.
\beq
V^{0K}\equiv h^K(V^{0I},e_{AK}V^A + S_K + \bar S_K)\ .
\eeq
 The precise relation will of course depend on the
particular form of $K(V^{0I})$. We may rewrite now $K$ in terms of
$h^K$ and replace it by its Legendre-transform $\hat K$\beq
\label{legendretrafomlos} - \hat{K}( e_{AI}V^A + S_I + \bar{S}_I )
= K(h^{J}) + ( e_{AI}V^A +
 S_I + \bar{S}_I)h^I \ ,
\eeq which, due to (\ref{legendrev}), is a function of $e_{AK}V^A
+ S_K + \bar S_K$.
Inserted into \eqref{firstordermasselos} we finally arrive at
 \beq
\label{dualemasselos} \cL =
\int d^4 \theta \big(\hat{K}(
e_{AI}V^A + (S_I + \bar{S}_I) ) + \tfrac14 \Big( \int d^2 \tx
f_{AB}W^AW^B + \textrm{ h.c. } \Big) \  . \eeq
$\cL$ is the dual
Lagrangian of \eqref{actionre} which is expressed in terms of $n_V$ vector multiplets $V^A$ and
$n_L$ chiral multiplets $S_I$.

In the original formulation given in \eqref{actionre} the gauge
invariance of the vector multiplets \beq \label{E-Tr-VSS-1} V^A
\to V^A + \Sx^A + \bar{\Sx}^A \ , \qquad \bar D_{\dot\alpha}
\Sigma^A =0\ , \eeq is manifest since the entire action is
expressed in terms of the gauge invariant field strength $W^A$. In
the dual formulation \eqref{E-Tr-VSS-1} has to be accompanied by a
shift of the chiral multiplets \beq \label{E-Tr-VSS-2} S_I \to S_I
- e_{AI}\Sx^A\ . \eeq We see that the $S_I$ play the role of
Goldstone supermultiplets which are necessary in order to maintain
the $U(1)$ gauge invariance. Thus the first term in
\eqref{dualemasselos} corresponds to a mass term for the vector
fields while the second term is the standard kinetic term. In
order to see this more explicitly let us expand the Lagrangian
\eqref{dualemasselos} in components. We take $V^A$ in a
Wess-Zumino gauge and expand accordingly \bea \label{KompVA} V^A
&=& -\tx\sx^m\tb v^A_m + i\tx\tx\tb\lb^A -i
\tb\tb\tx\lx^A + \tfrac12 \tx\tx\tb\tb D^A\ ,\nonumber \\
S_I &=& \tfrac12 E_I +\sqrt 2  \tx\psi_I + \tx\tx F_I\ . \eea
Inserted into \eqref{dualemasselos} we arrive at \bea \label{Wint}
{\cL}& = & -
\tfrac{1}{4}\mathrm{Re}f_{AB}F^{A\,mn}F^B_{mn}+\tfrac{1}{8}\IM
f_{AB}\epsilon^{mnpr}F^A_{mn}F^B_{pr} +
\tfrac{1}{2}\mathrm{Re}f_{AB}D^AD^B \nn \\ &\phantom{=}&-
\tfrac{i}{2}\left(f_{AB}\lambda^A\sigma^m\partial_m\bar{\lambda}^B
- \bar f_{AB}\partial_m\lambda^A\sigma^m\bar{\lambda}^B\right)\ +
\tfrac{i}{2\sqrt{2}}\,\partial_i f_{AB}D^A\chi^i\lambda^B
-\tfrac{i}{2\sqrt{2}}\,\partial_{\bar i}\bar
f_{AB}D^A\bar{\chi}^{\bar i}\bar{\lambda}^B \nn
\\ &\phantom{=}& -\tfrac{1}{2\sqrt{2}}\,\partial_i
 f_{AB}{F}^A_{mn}\chi^i\sigma^{mn}\lambda^B
 -\tfrac{1}{2\sqrt{2}}\,\partial_{\bar i}
 \bar f_{AB}{F}^A_{mn}\bar{\chi}^{\bar i}\bar{\sigma}^{mn}
 \bar{\lambda}^B
 \nn \\ &\phantom{=}& - \tfrac{1}{4}F^i\partial_i f_{AB}\lambda^A\lambda^B
-\tfrac{1}{4}\bar F^{\bar i}\partial_{\bar i}\bar
f_{AB}\bar{\lambda}^A\bar{\lambda}^B  \nn
 \\ &\phantom{=}&
+
\tfrac{1}{8}\chi^i\chi^j\partial_i\partial_jf_{AB}\lambda^A\lambda^B
+ \tfrac{1}{8}\bar{\chi}^{\bar i}\bar{\chi}^{\bar j}\partial_{\bar
i}\partial_{\bar j}f_{AB}
\bar{\lambda}^A\bar{\lambda}^B \nn  \\
&\phantom{=}& +\tfrac12 \hat K_Ie_{AI}D^A  +\tfrac14 \hat
K_{IJKL}\psi^I\psi^J\bar\psi^K\bar\psi^L \nn \\ &\phantom{=}&
+\hat K_{IJ}\{ - \tfrac14 \partial^m(\RE E_I)\partial_m(\RE E_J) -
\tfrac14 (\partial^m(\IM E_I)+ e_{AI}v^{A\,m})
(\partial^m(\IM E_J)+ e_{BJ}v^{B\,m}) \nn \\
&\phantom{=}& \phantom{+\tfrac12 \hat K_{IJ}\{} +\tfrac{i}{\sqrt2}
e_{AI}(\psi_J\lx^A - \bar\psi_J\lb^A) -
\tfrac{i}2(\psi_J\sx^m\partial_m\bar\psi_I +
\bar\psi_J\bar\sx^m\partial_m\psi_I) + F_I \bar F_{J} \} \nn \\
&\phantom{=}& +\tfrac12 \hat K_{IJK} \{ -\psi_I\sx^m\bar\psi_J(
e_{AK}v^{A\,m} +  \partial^m(\IM E_K)) -(\psi_I\psi_J)\bar F_K -
(\bar\psi_I\bar\psi_J)F_K \} \ , \eea where we abbreviate $\hat
K_I= \partial_{\RE E_I}\hat K$. As promised we see that the real
scalars ($\IM E_K$) play the role of $n_L$ Goldstone bosons which
render the linear combinations $e_{AI}v^{A\,m}$ of the $n_L$
vector fields $v^{A\,m}$ massive.

Eliminating the auxiliary fields $F_I$ and $D^A$ by their
equations of motion we arrive at the following bosonic action \bea
\label{ldualbosonoFH} \cL_{b}=& -
&\tfrac{1}{4}\mathrm{Re}f_{AB}F^{A\,mn}F^B_{mn}+\tfrac{1}{8}\IM
f_{AB}\epsilon^{mnpr}F^A_{mn}F^B_{pr} \nn \\
 &-& \tfrac14 \hat K_{IJ}\left( \partial^m(\RE E_I)\partial_m(\RE E_J) +
  e_{AI}e_{BJ}v^{A}_mv^{B\,m} \right) - V \ ,
\eea where we have chosen the unitary gauge and absorbed $\IM E_K$
into a redefinition of $v^{A}_m$. The scalar potential is of the
standard $N=1$ form and given by \bea \label{skalpotdual} V &=&
\tfrac12 \RE f_{AB} D^AD^B = \tfrac18 (\RE f)^{-1CD}
e_{CI}e_{DJ}\hat K_I \hat K_J \ . \eea This potentials agrees with
the one given in (\ref{skalares-Potential-C}) for
$m^A_{\phantom{A}I}=0$ if one also identifies $\hat K_I = - C^I$.
For this substitution also the kinetic terms of the scalars $C^I$
and $(\RE E_I)$ agree. Indeed starting from $-\frac14 K_{IJ}
(\partial_m C^J)(\partial^m C^I)$ and using the above
identification we arrive at $- \tfrac14 \hat K_{IJ}(\partial^m\RE
E_I)(\partial_m\RE E_J)$, taking into account that due to
(\ref{legendretrafomlos}) we have $\partial_{\RE E_K}K_J =
-\dx^J_K$.

\subsection{Massive antisymmetric tensors}

Let us now turn on the couplings $m^A_{\phantom{A}I}$ and repeat
the analysis of the previous section. In this case we start from
the first-order Lagrangian
 \bea
\label{firstordermassiv} \cL_{first} &=& \int d^4 \tx \left\{ \U
({\widetilde{V}^I}) -\tfrac12 \widetilde{V}^I(D^{\ax}\Phi^I_\ax
+\bar D_{\dot\ax}\bar \Phi^{I\, \dot\ax})  \right\} + \cL_m \eea
where $\cL_m$ is given in \eqref{Lm-VL}. $\U $ is a real function
of the vector multiplets $\tilde V^I$ which will turn out to be
the Legendre transform of $K$.

Let us first show that from (\ref{firstordermassiv}) one can
derive the Lagrangian for $n_L$ massive linear multiplets as given
by the sum of (\ref{Lm-VL}) and (\ref{nL-Lkin}). To do so we vary
(\ref{firstordermassiv}) with respect to $\widetilde{V}^J$ and
obtain \bea \label{lrevmass}
\partial_{\widetilde{V}^J}\U (\widetilde{V}^I) = \tfrac12 (D^{\ax}\Phi^J_\ax
+\bar D_{\dot\ax}\bar \Phi^{J\, \dot\ax}) = L^J  \ . \eea For
appropriate $\U $ (\ref{lrevmass}) may be solved giving
$\widetilde{V}^J$ as a function of $L^J$ and $\widetilde{V}^I$, $I
\neq J$.  We shall denote this function by $h^K=h^K(L^K,
\widetilde V^I)\equiv \widetilde V^{K}$. As in the massless case
we can express $\U $ in terms of the $h^K$. Together with
(\ref{lrevmass}) this leads us from (\ref{firstordermassiv}) to
\beq \label{lklmdualzw1} \cL = \int d^4 \tx \left\{ \U (h^K) -
h^IL^I \right\} + \cL_m \ . \eeq Due to (\ref{lrevmass}) the
expression $-K(L^I) := \U (h^I) - h^JL^J$ is the Legendre
transform of $\U (h^I)$ with respect to all $h^I$, i.e. a function
depending only on the $L^I$. Substituting $K(L^I)$ into
(\ref{lklmdualzw1}) we have arrived at the Lagrangian for $n_L$
massive linear multiplets as stated above.

Alternatively we can eliminate the $\Phi^I_\ax$ multiplets and
this yields the desired dual action. To do so let us first rewrite
(\ref{firstordermassiv}) as \bea \label{firstorder-P} \cL_{first}
&=&  \int d^4 \tx \U ({\widetilde{V}^I}) + \tfrac{1}{4}\int d^2\tx
f_{AB}W^AW^B + \rm{ h.c.}\\ &\phantom{=}&+  \int d^2 \tx \Big\{
\Phi^I \left(\tfrac12 \widetilde{W}^I + \tfrac12 e_{AI} W^A
-if_{AB}m^B_{\phantom{B}I}W^A\right) - \M^2_{IJ} \Phi^I \Phi^J
\Big\}  + \rm{ h.c.} \ ,  \nn \eea where $\widetilde{W}^J =
-\tfrac14 \bar D^2 D \widetilde{V}^J$ is the field strength of
$\widetilde{V}^J$ and we have performed a partial integration. We
also introduced the mass matrix \beq \label{defMij} \M^2_{IJ}:=
(M^2)_{IJ} + i (M^2_T)_{IJ} \ , \eeq with $M^2$ and $M^2_T$ being
defined in (\ref{M2undM2T}). The equation of motion for $\Phi_\ax$
can be obtained from
 (\ref{firstorder-P}) by using again $\Phi_\ax= \bar D^2 X_\ax$.
Demanding $\M^2_{IJ}$ to be invertible we arrive at \beq
\label{bewglnPhii} \Phi^I_\ax = \tfrac12 (\M^2)^{-1}_{IK}(\tfrac12
\widetilde{W}^K_\ax + \tfrac12 e_{AK}W^A_\ax
-if_{AB}m^B_{\phantom{B}K}W^A_\ax) \ . \eeq Inserting back into
(\ref{firstordermassiv}) we obtain \beq \label{Ldualmassivfhat}\cL
= \int d^4 \tx \U ({\widetilde{V}^i}) + \tfrac{1}{4}\int d^2\tx
\hat f_{\hat A \hat B}W^{\hat A}W^{\hat B} + \rm{ h.c.} \ ,  \eeq
where we have introduced $ W^{\hat A}:= \big( -\tfrac12
\widetilde{W}^I, W^A \big)$. So the index  $\hat A$ takes values
$\hat A= (I, A)=(1, \dots, n_L, n_L+1, \dots,n_L+n_V )$.
Furthermore the $(n_V+n_L)\times(n_V+n_L)$ - dimensional gauge
coupling matrix $\hat f_{\hat A \hat B}$ is given by \beq
\label{fhat} \hat f_{\hat{A}\hat{B}} = \left(
\begin{array}{cc} \hat f_{IJ} & \hat f_{IB} \\ \hat f_{AJ} & \hat
  f_{AB} \end{array}\right)\ ,
\eeq where
\begin{equation} \begin{split}\label{hat-fAB} \hat f_{IJ} =&
(\M^2)^{-1}_{IJ} \ , \qquad \hat f_{IA} = (\M^2)^{-1}_{IK}\left(
-\tfrac12e_{Ak} + i f_{AD} m^D_{\phantom{D}K} \right) \\
 \hat f_{AB}=& f_{AB} + (\M^2)^{-1}_{IJ}\left(
-\tfrac12 e_{AI} + i f_{AD} m^D_{\phantom{D}I} \right)\left(
-\tfrac12 e_{BJ} + i f_{BC} m^C_{\phantom{C}J} \right) \
.\end{split}
\end{equation}

The term $\U ({\widetilde{V}^I})$ in the  Lagrangian
(\ref{Ldualmassivfhat}) contains mass terms for
$n_L$ vector multiplets. Thus the  Lagrangian
(\ref{Ldualmassivfhat}) appears to depend
on $n_V$ massless and $n_L$ massive vector multiplets.
However $n_L$ of the original $n_V$ vector fields are now
unphysical. This can be seen from the fact that the gauge coupling
matrix $\RE \hat f_{\hat A \hat B}$ has $n_L$ zero eigenvalues
while $\IM \hat f_{\hat A \hat B}$ has $n_L$ constant
eigenvalues. Or in other words
$n_L$ of the vector fields only have a topological coupling but no
kinetic term. Indeed using (\ref{M2undM2T}) and (\ref{defMij}) it is easy
to verify that
\beq
\hat f_{IB}\, m^B_{\phantom{B}K} = i \dx_{IK},
\qquad  \hat f_{AB}\, m^B_{\phantom{B}K} = -\tfrac{i}2\, e_{AK} \ .
\eeq
This shows that the $n_L$ vectors
$(0, \dots, 0, m^B_{\phantom{B}K})$ are eigenvectors of
$\RE \hat f_{\hat A \hat B}$ with eigenvalue zero.

In order to display the physical components of $\widetilde V$ we
decompose it into a vector multiplet $V^{0I}$ in the WZ-gauge and
the real part of a chiral superfield $S^I$ \beq \label{V-tilde}
\widetilde{V}^I := V^{0I}+S^I+ \bar S^I \ . \eeq The component
form of (\ref{Ldualmassivfhat}) can then be obtained by inserting
the  Wess-Zumino gauge \eqref{KompVA} for $V^{0I}$ while for the
chiral multiplets $S^I$ we use \bea S_I &=& \tfrac12 A_I + \sqrt2
\tx \psi_I + \tfrac{i}{2} \tx\sx^m\tb \partial_m A_I + \tx\tx F_I
- \tfrac{i}{\sqrt2} \tx \tx
\partial_m \psi_I \sx^m \tb + \tfrac14 \tx\tx\tb\tb \Box \tfrac12 A_I \ .
\eea
Inserted into (\ref{Ldualmassivfhat}) we arrive at
\bea \label{KompLdualmassiv} \cL &=& -\tfrac14 \RE
\hat{f}_{\hat{A}\hat{B}}F^{\hat{A}\,mn}F^{\hat{B}}_{mn} +\tfrac18
\IM \hat{f}_{\hat{A}\hat{B}}
\varepsilon^{klmn}F^{\hat{A}}_{kl}F^{\hat{B}}_{mn} +\tfrac12 \RE
\hat{f}_{\hat{A}\hat{B}}D^{\hat{A}}D^{\hat{B}} \nn \\
&\phantom{=}& + \tfrac12 \U _ID^{0I} - \tfrac14 \U _{IJ}
v^{0I\,m}v^{0J}_m - \tfrac14 \U _{IJ}\partial^m(\RE
A_I)\partial_m(\RE A_J) - \U _{IJ}F_I\bar F_{J}\nn\\ &\phantom{=}&
- \tfrac{i}{2}\left( \hat{f}_{\hat{A}\hat{B}} \lx^{\hat{A}}
\sx^k\partial_k \bar\lx^{\hat{B}} + \bar{\hat{f}}_{\hat{A}\hat{B}}
\bar\lx^{\hat{A}} \bar\sx^k\partial_k \lx^{\hat{B}} \right)\nn \\
&\phantom{=}& +\tfrac12 \U _{IJ} \left\{ i\sqrt 2
(\psi_J\lx^{0I}-\bar\psi_J\lb^{0I})-i(\psi_J\sx^m\partial_m\bar\psi_I
+ \bar\psi_J\bar\sx^m\partial_m\psi_I ) \right\}\nn \\
&\phantom{=}& +\tfrac12 \U _{IJK}\left\{ - \psi_I\sx^m\bar\psi_J
v^{0K}_m -(\psi_I\psi_J)\bar F_K - (\bar\psi_I \bar\psi_J)F_K
\right\} \nn
\\ &\phantom{=}& +\tfrac14
\U _{IJKL}\psi_I\psi_J\bar\psi_K\bar\psi_L + \dots \ , \eea where
$\U _I= \partial_{\RE A_I} \U$ was used and terms proportional to
$\partial_i \hat{f}_{\hat{A}\hat{B}}$ have been neglected.

The next step is to eliminate the auxiliary fields from the
Lagrangian. The equations of motion for $F_I$ can be determined in
a straightforward manner to be
\bea \label{eom-Fi}
F_I = \tfrac12
\U ^{-1}_{IL}\U _{LJK} \psi_J\psi_K \ .
\eea
For $D^{\hat A}$ however
the situation is more difficult since some of the vector
multiplets are unphysical.
In order to remove the unphysical degrees of freedom
we fix the gauge invariance of (\ref{bewglnPhii}).

To this aim let us rewrite (\ref{bewglnPhii}) in the following way
\bea \label{PhiREIM} \Phi^I_\bx &=& -\tfrac12 R_{IK} \Big\{ -
\tfrac12 \widetilde{W}^K - \left( \tfrac12 e_{AK} + \IM f_{AB}
m^B_{\phantom{B}K}\right) W^A  + R^{-1}_{KL} I_{LJ} \RE f_{AB}
m^B_{\phantom{B}J}W^A  \Big\}\nn \\
 &+&
\tfrac i2 I_{IK} \Big\{ \tfrac12 \widetilde{W}^{K} + \left(
\tfrac12 e_{AK} + \IM f_{AB} m^B_{\phantom{B}K}\right)  W^A  -
I^{-1}_{KL} R_{LJ} \RE f_{AB} m^B_{\phantom{B}J}W^A \Big\}  , \eea
where we divided the coupling matrices of (\ref{bewglnPhii}) into
their real and imaginary parts and abbreviated \beq R_{IK} = [\RE
((\M^2)^{-1})]_{IK} \qquad \textrm{ and } \qquad I_{IK} = [\IM
((\M^2)^{-1})]_{IK} \ . \eeq Going to the WZ-gauge
\eqref{Phiexpansion1} for $\Phi^I$ we see that the
$\theta$-component of the imaginary part of the right hand side of
\eqref{PhiREIM} has to vanish. This implies \beq
\label{transfEichbed}
 D^{0K} =
- \left( e_{AK}  + 2 \IM f_{AB} m^B_{\phantom{B}K}\right) D^A + 2
I^{-1}_{KL}  R_{LJ} \RE f_{AB}
 m^B_{\phantom{B}J}D^A + \ldots\ ,
\eeq
where we have omitted the fermionic terms.
We can now use the constraint \eqref{transfEichbed}
to eliminate the $D^{0K}$ from
the Lagrangian. Let us concentrate on the bosonic terms which we
read off from (\ref{KompLdualmassiv}) to be
\beq \label{skalpotmassiv}
V = -\tfrac12 \RE
\hat{f}_{\hat{A}\hat{B}}D^{\hat{A}}D^{\hat{B}} - \tfrac12
\U _I D^{0I} \ . \eeq
 Inserting (\ref{hat-fAB}), (\ref{transfEichbed}) and using $D^{\hat
A} = (-\tfrac12 D^{0K}, D^A)$ we obtain the following equation of
motion for $D^A$ {\setlength\arraycolsep{0pt}\bea \label{kpm1}
&{}& \left\{ R_{IJ} I^{-1}_{JK} I^{-1}_{IN} R_{KL} R_{NS} \RE
f_{AC} \RE f_{BD} m^C_{\phantom{C}L} m^D_{\phantom{D}S} + R_{IJ}
\RE f_{AC} \RE f_{BD} m^C_{\phantom{C}I}
m^D_{\phantom{D}J} + \RE f_{AB} \right\} D^B \nn \\
&{}& \phantom{=} = - \U _K \left\{ - \left( \tfrac12 e_{AK} + \IM
f_{AC} m^C_{\phantom{C}K}\right) + I^{-1}_{KL} R_{LJ} \RE f_{AC}
m^C_{\phantom{C}J} \right\} \ . \eea} The inverse of the matrix
multiplying $D^B$ is found to be \bea \label{invkpm} (\RE
f)^{-1EA} - R_{TU}m^E_{\phantom{E}T}m^A_{\phantom{A}U} \ , \eea
which implies \bea \label{eom-DA-dual} D^E =  \U _K (\RE f)^{-1EA}
\left( \tfrac12 e_{AK} + \IM f_{AC}m^C_{\phantom{C}K} \right) \ .
\eea Inserting (\ref{eom-Fi}) and (\ref{eom-DA-dual}) back into
(\ref{KompLdualmassiv}) we arrive at
 \bea \label{dualfinal} \cL
&=& -\tfrac14 \RE
\hat{f}_{\hat{A}\hat{B}}F^{\hat{A}\,mn}F^{\hat{B}}_{mn} +\tfrac18
\IM \hat{f}_{\hat{A}\hat{B}}
\varepsilon^{klmn}F^{\hat{A}}_{kl}F^{\hat{B}}_{mn}  - \tfrac14 \U
_{IJ} v^{0I\,m}v^{0J}_m \nn \\ &\phantom{=}& - \tfrac14 \U
_{IJ}\partial^m(\RE A_I)\partial_m(\RE A_J) - \tfrac{i}{2}\left(
\hat{f}_{\hat{A}\hat{B}} \lx^{\hat{A}} \sx^k\partial_k
\bar\lx^{\hat{B}} + \bar{\hat{f}}_{\hat{A}\hat{B}}
\bar\lx^{\hat{A}} \bar\sx^k\partial_k \lx^{\hat{B}} \right)\nn \\
&\phantom{=}& +\tfrac12 \U _{IJ} \left\{ i\sqrt 2
(\psi_J\lx^{0I}-\bar\psi_J\lb^{0I})-i(\psi_J\sx^m\partial_m\bar\psi_I
+ \bar\psi_J\bar\sx^m\partial_m\psi_I ) \right\}\nn \\
&\phantom{=}& -\tfrac12 \U _{IJK} \psi_I\sx^m\bar\psi_J v^{0K}_m
\nn \\ &\phantom{=}& +\tfrac14 \left( \U _{IJKL} - \U _{IJM}\U
^{-1}_{MS}\U _{KLS} \right) \psi_I\psi_J\bar\psi_K\bar\psi_L -V  +
\dots \ , \eea where the scalar potential is given by \beq
\label{skalpotnachD} V = \tfrac{1}{8} \left\{ \left( e_{AI} + 2
\IM f_{AC}m^C_{\phantom{C}I} \right)\RE f^{-1AB} \left( e_{BJ} + 2
\IM f_{BD}m^D_{\phantom{D}J} \right) + 4 \RE
f_{AB}m^A_{\phantom{A}I}m^B_{\phantom{B}J} \right\} \U _I \U _J \
. \eeq This potential indeed coincides with
(\ref{skalares-Potential-C}) for $C^I =  \U ^I$. For this
identification also the kinetic terms agree which is expressing
simply the fact that $K$ and $\U $ are related to each other by a
Legendre transformation. Thus \eqref{dualfinal} is the desired
dual action of \eqref{lmkompohneHF}.


\section{Conclusion}

Let us summarize our results. We proposed an $N=1$ superfield
action for $n_L$  chiral spinor superfields
coupled to $n_V$ vector and $n_C$ chiral multiplets.
The component form of this action was given and shown to contain gauge
invariant mass terms for  $n_L$ antisymmetric tensors. In addition the
action also features
Green-Schwarz couplings to the $n_V$ vector fields.
Supersymmetry gives a mass to the supersymmetric partners $C^I$
of the antisymmetric tensors with the peculiarity that these  mass
terms  do not arise from eliminating an auxiliary field. Indeed the
supersymmetry transformation laws show that any Lorentz invariant ground
state of the spinor superfield preserves supersymmetry. Instead the
supersymmetry transformations of the vector multiplets are modified
and a vacuum expectation value of the scalars $C^I$ can break
supersymmetry by generating a non-vanishing gaugino transformation.

 We also constructed the dual action in terms of
$n_L$ massive and $n_V-n_L$ massless vector
multiplets by explicitly
performing the duality transformations in
superspace.  We gave the component form of the dual action and
showed that the scalar potentials in both formulations
coincide.

For one chiral spinor superfield the
action agrees with the action given in \cite{Louis-W} which also
appeared in Kaluza-Klein reduction of
type IIB string theory compactified on Calabi-Yau orientifolds
\cite{Grimm-Louis-IIB}.
\vskip 1cm


\subsection*{Acknowledgments}

This work is supported by The German Science Foundation (DFG) and the
European RTN Program  MRTN-CT-2004-503369.

We thank  Ferdinand Brennecke, Olaf Hohm, Waldemar Schulgin and Ulrich Theis
for useful discussions.

\appendix
\section{Modified SUSY-transformations of the chiral spinor superfield}

In this appendix we derive the supersymmetry transformation laws
of a chiral spinor superfield in the WZ-gauge. The main motivation for
this exercise
is to identify the order parameters for spontaneous supersymmetry
breaking.
For simplicity we perform this analysis for a single  $\Phi_\ax$.

The general supersymmetry transformation of $\Phi_\ax$ reads \beq
\Phi_\ax \to  \Phi'_\ax = \Phi_\ax +\dx_\xi \Phi_\ax = \Phi_\ax +
(\xi Q + \bar \xi \bar Q) \Phi_\ax \ , \eeq where $Q$ and $\bar Q$
are the supersymmetry generators. In terms of the component
expansion \eqref{entphi} we have \beq \label{susytrafospinor}
\begin{split}
\delta_{\xi} \chi_{\ax} = & -\xi_{\gamma}\Big(\tfrac12\,
{\delta_{\alpha}}^{\gamma}(C+iE) +
\tfrac{1}{4}{(\sigma^m\bar{\sigma}^n)_{\alpha}}^{\gamma}B_{mn}\Big)
\ ,
\\ \dx_\xi C = & \xi^{\ax}\eta_{\ax} +
\bar{\xi}_{\dot{\ax}}\bar{\eta}^{\dot{\ax}} \ , \\ \dx_\xi E = &
-i\xi^{\ax}\eta_{\ax} +
i\bar{\xi}_{\dot{\ax}}\bar{\eta}^{\dot{\ax}} + 2\left(
\xi^{\beta}\sx^m_{\bx\dot{\ax}}\partial_m \bar{\chi}^{\dot{\ax}} -
\bar{\xi}_{\dot{\ax}}\bar{\sx}^{m\dot{\ax}\bx}\partial_m\chi_{\bx}
\right) \ , \\ \dx_\xi B^{mn} = & 2
\eta^{\ax}{(\sx^{mn})_{\ax}}^{\bx}\xi_{\bx} +  2
\bar{\eta}_{\dot{\ax}}{(\bar{\sx}^{mn})^{\dot{\ax}}}_{\dot{\bx}}\bar{\xi}^{\dot{\bx}}
\\
 & + 2i (  \xi^{\beta}\sx^m_{\bx\dot{\ax}}\partial^n \bar{\chi}^{\dot{\ax}} -
  \xi^{\beta}\sx^n_{\bx\dot{\ax}}\partial^m \bar{\chi}^{\dot{\ax}} ) + 2i
  (\bar{\xi}_{\dot{\ax}}\bar{\sx}^{m\dot{\ax}\bx}\partial^n\chi_{\bx}
-
\bar{\xi}_{\dot{\ax}}\bar{\sx}^{n\dot{\ax}\bx}\partial^m\chi_{\bx})
\ ,
\\ \dx_\xi \eta_{\ax} = &
i\sx^k_{\ax\dot{\ax}}\bar{\xi}^{\dot{\ax}}\partial_kC - \tfrac12
\varepsilon^{kmnr}\sx_{r\ax\dot{\ax}}\bar{\xi}^{\dot{\ax}}\partial_kB_{mn}
\ .
\end{split}
\eeq In the WZ-gauge we choose $\chi_{\ax} = 0$ and $E=0$ and thus
(\ref{susytrafospinor}) becomes \beq \label{susytrafosspinorwz}
\begin{split}
\delta_{\xi,WZ} \chi_{\ax} = &
-\xi_{\gamma}\Big(\tfrac12\,{\delta_{\alpha}}^{\gamma}C +
\tfrac{1}{4}{(\sigma^m\bar{\sigma}^n)_{\alpha}}^{\gamma}B_{mn}\Big)
\ ,
\\ \delta_{\xi,WZ} C = & \xi\eta + \bar{\xi}\bar{\eta} \ , \\ \delta_{\xi,WZ} E
= & -i\xi\eta + i\bar{\xi}\bar{\eta} \ , \\ \delta_{\xi,WZ} B^{mn}
=
& 2 \eta\sx^{mn}\xi +  2 \bar{\eta}\bar{\sx}^{mn}\bar{\xi} \ , \\
\delta_{\xi,WZ} \eta_{\ax} = & i(\sx^k\bar{\xi})_{\ax}\partial_kC
- \tfrac12
\varepsilon^{kmnr}(\sx_r\bar{\xi})_{\ax}\partial_kB_{mn} \ .
\end{split} \eeq
We see that $\chi_{\ax}$ and $E$ do not transform to zero and
therefore one needs a compensating gauge transformation to stay in
the WZ-gauge. These are given in (\ref{nL-eichtrPhi}) and
(\ref{eichtermPhi}) and so we are led to choose
\beq \label{eichwahl}
\begin{split}
\lx_\ax^e = &
-\xi_{\gamma}\Big(\tfrac12\, {\delta_{\alpha}}^{\gamma}C +
\tfrac{1}{4}{(\sigma^m\bar{\sigma}^n)_{\alpha}}^{\gamma}B_{mn}\Big)\ ,
\\ D^e = & i\xi\eta -i\bar{\xi}\bar\eta \ .
\end{split}
\eeq
This ensures $(\delta_{\xi, WZ} + \delta_{gauge})\chi = 0 =
(\delta_{\xi, WZ} + \delta_{gauge})E$ and modifies
the transformations of the physical
fields according to
 \beq \label{susyeichwzspinor}
\begin{split}
\left( \delta_{\xi, WZ} + \delta_{gauge} \right) C = &
\xi\eta + \bar\xi\bar\eta\ , \\ \left( \delta_{\xi, WZ} +
\delta_{gauge} \right) B^{mn} = & 2\eta\sx^{mn}\xi + 2
\bar\xi\bar\sx^{nm}\bar\eta +
\partial^m\Lx^{en} - \partial^n\Lx^{em}\ , \\
\left( \delta_{\xi, WZ} + \delta_{gauge} \right) \eta_{\ax}
= & i(\sx^k\bar{\xi})_{\ax}\partial_kC - \tfrac12
\varepsilon^{krmn}(\sx_r\bar{\xi})_{\ax}\partial_kB_{mn}  \ .
\end{split}
\eeq We see that in a Lorentz invariant ground state supersymmetry
cannot be broken by any of these transformations. However, in a
WZ-gauge
 the transformation laws of the charged multiplets also change.
For the case at hand these are the transformations of the vector multiplet
which without couplings to a spinor superfield read
\beq \label{susytrafovektor}
\begin{split}
\delta_\xi F_{mn} = & i\left[\left( \xi\sx^n\partial_m\lb +
\bar{\xi}\bar{\sx}^n\partial_m\lx \right) - \left(
\xi\sx^m\partial_n\lb + \bar{\xi}\bar{\sx}^m\partial_n\lx \right)
\right] \ , \\ \delta_\xi \lx_\ax = & i\xi_\ax D +
(\sx^{mn}\xi)_\ax F_{mn} \ ,  \\ \delta_\xi D = &
\bar\xi\bar\sx^m\partial_m\lx - \xi\sx^m\partial_m\lb \ .
\end{split}
\eeq
Gauge invariance of the couplings to the spinor superfield forces the
gauge fields to transform according to \eqref{trsuplang}. Thus
with the special choice (\ref{eichwahl}) we obtain for the
combined supersymmetry and gauge transformations
 \beq \label{susyeichwzvektor}
\begin{split}
\left( \delta_\xi + \delta_{gauge} \right) \lx_\ax = &
-m\xi_\gx\left( {\delta_\ax}^\gx C + \tfrac12 {(\sx^m
\bar{\sx}^n)_\ax}^\gx B_{mn} \right) + i\xi_\ax D +
(\sx^{mn}\xi)_\ax F_{mn} \ , \\ \left( \delta_\xi + \delta_{gauge}
\right)  D = & m(i\xi\eta - i\bar\xi\bar\eta) +
\bar\xi\bar\sx^m\partial_m\lx - \xi\sx^m\partial_m\lb \ , \\
\left( \delta_\xi + \delta_{gauge} \right)  F_{mn} = &
i\left[\left( \xi\sx^n\partial_m\lb +
\bar{\xi}\bar{\sx}^n\partial_m\lx \right) - \left(
\xi\sx^m\partial_n\lb + \bar{\xi}\bar{\sx}^m\partial_n\lx \right)
\right] +mF_{mn}^e \ .
\end{split}
\eeq As one can see supersymmetry can be broken in
(\ref{susyeichwzvektor}) if either $C$ or $D$ acquire a vacuum
expectation value that is different than zero.

\end{document}